%% file: manuscript.tex
\documentclass[runningheads]{llncs}

\input{packages.tex}

\begin{document}

\title{Structural invariants in individuals language use: the ``ego network" of words}

\author{Kilian Ollivier\inst{1}\orcidID{0000-0003-2881-5845} (\Letter) \and
Chiara Boldrini\inst{1}\orcidID{0000-0001-5080-8110} \and
Andrea Passarella\inst{1}\orcidID{0000-0002-1694-612X} \and
Marco Conti\inst{1}\orcidID{0000-0003-4097-4064}}

\authorrunning{Ollivier et al.}

\institute{CNR-IIT, Via G. Moruzzi 1, 56124, Pisa, Italy\\
\email{\{kilian.ollivier,chiara.boldrini,andrea.passarella,marco.conti\}@iit.cnr.it}}

\maketitle

\begin{abstract}

\input{abstract.tex}

\end{abstract}

\section{Introduction}

\label{sec:intro}

\input{introduction.tex}
\section{The dataset}
\label{sec:dataset}

\input{dataset.tex}
\vspace{-10pt}
\section{From word usage to cognitive constraints}
\label{sec:body}

\input{body.tex}
\vspace{-20pt}
\section{Conclusion}
\label{sec:conclusion}

\input{conclusion.tex}

\subsubsection*{Acknowledgements.}

\input{acks.tex}


\appendix

\section{Appendix}
\label{sec:appendix}

\input{appendix.tex}

\bibliographystyle{splncs04}  
\bibliography{bibliography}  

\end{document}

%% file: packages.tex
\usepackage{graphicx} 
\usepackage{soul} 
\usepackage{booktabs}
\usepackage{url}

\usepackage[strings]{underscore}

\usepackage{hyperref}
\usepackage{amsfonts}
\usepackage{nicefrac}
\usepackage{microtype}
\usepackage{lipsum}
\usepackage{tabularx}
\usepackage{caption} 
\usepackage{float}
\usepackage{subcaption}
\usepackage{caption}
\usepackage{dcolumn}
\usepackage{booktabs}
\usepackage{adjustbox}

\usepackage{multirow}
\usepackage[table,xcdraw]{xcolor}

\usepackage{marvosym}


%% file: abstract.tex

The cognitive constraints that humans exhibit in their social interactions have been extensively studied by anthropologists, who have highlighted their regularities across different types of social networks. We postulate that similar regularities can be found in other cognitive processes, such as those involving language production. In order to provide preliminary evidence for this claim, we analyse a dataset containing tweets of a heterogeneous group of Twitter users (regular users and professional writers). Leveraging a methodology similar to the one used to uncover the well-established social cognitive constraints, we find that a  concentric layered structure (which we call \emph{ego network of words}, in analogy to the ego network of social relationships) very well captures how individuals organise the words they use. The size of the layers in this structure regularly grows (approximately 2-3 times with respect to the previous one) when moving outwards, and the two penultimate external layers consistently account for approximately 60\% and 30\% of the used words (the outermost layer contains 100\% of the words), irrespective of the number of the total number of layers of the user.

%% file: introduction.tex
\vspace{-10pt}

Language production relies on many cognitive processes that unfold at every sequential step of word retrieval, essentially during the selection of the lexical element that will symbolize the concept that has to be expressed in the sentence~\cite{levelt1999theory}. The brain has acquired strategies to process efficiently the mental lexicon that contains a large number of words (e.g., for retrieving a single word among 40K others in less than 250ms). These strategies are executed unconsciously and they take advantage of language properties such as word frequencies~\cite{broadbent1967word,qu2016tracking} (the brain takes less time to retrieve words that are commonly used) to spend the least amount of time and effort in this task. In this work, we are interested in finding indirect traces of these cognitive processes in the written production. 
%
%
%
Specifically, we aim at investigating, through a data-driven approach, whether a regular structure can be found in the way people use words, as a ``symptom'' of cognitive constraints in this mental process. We argue that words usage might present similar properties to other mental processes which are known to be driven by cognitive constraints, specifically the way how humans allocate cognitive capacity to maintaining social relationships.

\begin{figure}[t]
\begin{center}
\includegraphics[scale=0.3]{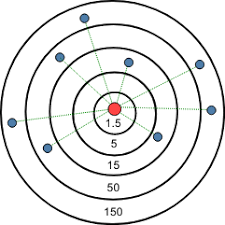}\vspace{-5pt}
\caption{The \emph{ego network} of social relationships. The red dot symbolizes the ego and the blue dots the alters with whom the ego maintains an active social relationship. The numbers correspond to the layers sizes.}
\label{fig:egonet}\vspace{-30pt}
\end{center}
\end{figure}

The cognitive efforts that we allocate to socialization have been extensively studied by anthropologists, and their findings~\cite{dunbar1998social} show that the social life of humans is constrained, through time and cognitive capacity, to 150 meaningful relationships per person (a limit that goes under the name of \emph{Dunbar's number}, from the scientist who first postulated its existence). This limit is observable in primates as well, where it is related to how many peers can be effectively groomed by the animals to reinforce social bonds. In humans, these 150 social relationships can be grouped into classes of different intimacy. Specifically, anthropologists have found that the social relations around the average individual can be grouped into at least 4 concentric layers~\cite{hill2003social,Zhou2005}, starting from the innermost one which typically includes our closest family members. The typical sizes of these layers are 5, 15, 50, 150. Many people also feature an additional layer, whose size is around 1.5 people, included in the first layer, which comprises relationships with extremely high social closeness~\cite{dunbar2015structure}. This structure of social relationships, illustrated in Figure~\ref{fig:egonet}, is typically referred to as \emph{ego network}. 
A characteristic fingerprint of these social circles is their scaling ratio (i.e. the ratio between the sizes of consecutive layers), which has been found to be approximately around 3, regardless of the specific social network considered. Interestingly, both real-life and online social networks follow this social organization~\cite{dunbar2015structure,Haerter2012,Miritello2013,gonccalves2011modeling}.
The discovery of this social structure (stratified in concentric layers) and its invariants (in terms of number of layers and scaling ratio across different and heterogeneous social networks) has represented a breakthrough moment in this research area. Many subsequent studies have leveraged this aggregate representation through social circles to better understand social-dependent human behaviour, such as how humans trust each other~\cite{sutcliffe2015modelling} or how they share resources and information~\cite{aral2011diversity,arnaboldi2014information}. 

Building upon the above considerations, in this work we set out to investigate the presence of an analogous structure and structural invariants in cognitive processes beyond the well-established social ones. Indeed, socialization is just one of the many cognitive processes that we entertain in our daily life. Thus, it is reasonable to expect that similar limitations in our cognitive capacity yield characteristic structural properties in other domains as well. Here, we focus on the cognitive process associated with language production, which, as described above, is tightly related to our cognitive capacity. Moreover, language is intimately linked to sociality, as there are hypotheses (that go under the name of \emph{social gossip theory of language evolution}~\cite{Dunbar1998}) postulating that language has been developed as a more efficient way for grooming social relationships: with \emph{vocal grooming}, we can reach more peers at the same time. We already had pieces of evidence suggesting the existence of cognitive limits in language production. One of the most prominent examples, which has been reported by G. Zipf \cite{zipf1949human} in 1932, is the empirical observation that the frequency of words in a corpus is inversely proportional to its position in the frequency table. It is also well-known that our vocabulary size is limited: e.g., an average 20-year-old native speaker of American English knows 42,000 words~\cite{Brysbaert2016}.
In this work, we try to go one step further and investigate more complex structural properties, leveraging the approach used to uncover cognitive constraints in the social domain.
To the best of our knowledge, this research perspective has never been tackled before in the related literature. 
Please note that when we refer to the structural properties of language use, we do not refer to grammar but to the language-agnostic way in which lemmas are assigned a cognitive effort by their users.

An advantage of studying language production is that myriad textual datasets are available online. In trying to find out which cognitive constraints affect the production of language, our intuition is that the more spontaneous this production is, the greater the time constraint, and the more visible the cognitive limit. We argue that Twitter is a platform that facilitates a spontaneous writing style, much more so than newspaper articles or speech transcripts (just to mention a few other textual formats readily available online). For this reason, we choose Twitter for this initial investigation of the cognitive constraints in language production, leaving the analysis of other textual dataset as future work. 
%
We have collected a diverse dataset of tweets for our analysis, including tweets from regular Twitter users and professional writers (Section~\ref{sec:dataset}). Then, leveraging a methodology similar to the one used to uncover social constraints, we study the structural properties of language production on Twitter as a function of the individual word usage frequency, and we provide preliminary evidence for a set of cognitive constraints that naturally determine the way we communicate (Section~\ref{sec:body}). Specifically, our main findings are the following:
\vspace{-3pt}
\begin{itemize}
    \item Similarly to the social case, we found that a \emph{regular concentric, layered structure} (which we call \emph{ego network of words} in analogy to the ego networks of the social domain) very well captures how an individual organizes their cognitive effort in language production. Specifically, words can be typically grouped in between 5 and 7 layers of decreasing usage frequency moving outwards, regardless of the specific class of users (regular vs professional) and of the specific time window considered.
    \item One structural invariant is observed for the \emph{size of the layers}, which approximately doubles when moving from layer $i$ to layer $i+1$. The only exception is the innermost layer, which tends to be approximately 5 five times smaller than the next one. This suggests that the innermost layer, the one containing the most used words, may be drastically different from the others.
    \item A second structural invariant emerges for the \emph{external layers}. Users with more clusters organise differently their innermost layers, without modifying significantly the size of the most external ones. In fact, while the size of all layers beyond the first one linearly increases with the most external layer size, the second-last and third-last layer consistently account for approximately 60\% and 30\% of the used words, irrespective of the number of clusters of the user.
\end{itemize}

\vspace{-15pt}

%% file: dataset.tex

\vspace{-10pt}

The analysis is built upon four datasets extracted from Twitter, using the official Search and Streaming APIs (note that the number of downloadable tweets is limited to 3200 per user). Each of them is based on the tweets issued by users in four distinct groups:
\begin{description}
\item[Journalists] Extracted from a Twitter list containing New York Times journalists\footnote{\url{https://twitter.com/i/lists/54340435}}, created by the New York Times itself. It includes 678 accounts, whose timelines have been downloaded on February 16th 2018. 
\item[Science writers] Extracted from a Twitter list created by Jennifer Frazer\footnote{\url{https://twitter.com/i/lists/52528869}}, a science writer at \textit{Scientific American}. The group is composed of 497 accounts and has been downloaded  on June 20th 2018.
\item[Random users \#1] This group has been collected by sampling among the accounts that issued a tweet or a retweet in English with the hashtag \textit{\#MondayMotivation} (at the download time, on January 16th 2020). This hashtag is chosen in order to obtain a diversified sample of users: it is broadly used and does not refer to a specific event or a political issue. As the accounts are not handpicked as in the two first groups, we need to make sure that they represent real humans. The probability that an account is a bot is calculated with the Botometer service~\cite{davis2016botornot}, which is based not only on language-agnostic features like the number of followers or the tweeting frequency, but also on linguistic features such as grammatical tags, or the number of words in a tweet~\cite{varol2018feature}. The algorithm detects 29\% of bot accounts, such that this dataset is composed of 5183 users.
\item[Random users \#2] This group has been collected by sampling among the accounts  which issued a tweet or a retweet in English, from the United Kingdom (we set up a filter based on the language and country), at download time on February 11th 2020. 23\% of the accounts are detected as bot, such that this group contains 2733 accounts. 
\end{description} \vspace{-5pt}
These groups are chosen to cover different types of users: the first two contain accounts that use language professionally (journalists and science writers), the other two contain regular users, which are expected to be more colloquial and less controlled in their language use.
Please note that we discard retweets with no associated comments, as they do not include any text written by the target user, and tweets written in a language other than English (since most of the NLP tools needed for our analysis are optimised for the English language).
In our analysis, we only consider active Twitter accounts, which we define as an account not abandoned by its user and that tweets regularly. Further details on this preprocessing step are provided in Appendix~\ref{app:activeusers}.


\subsection{Extracting user timelines with the same observation period}
\vspace{-8pt}

\begin{figure}[t]
\centering
\begin{minipage}{.48\textwidth}
  \centering
  \includegraphics[width=\linewidth]{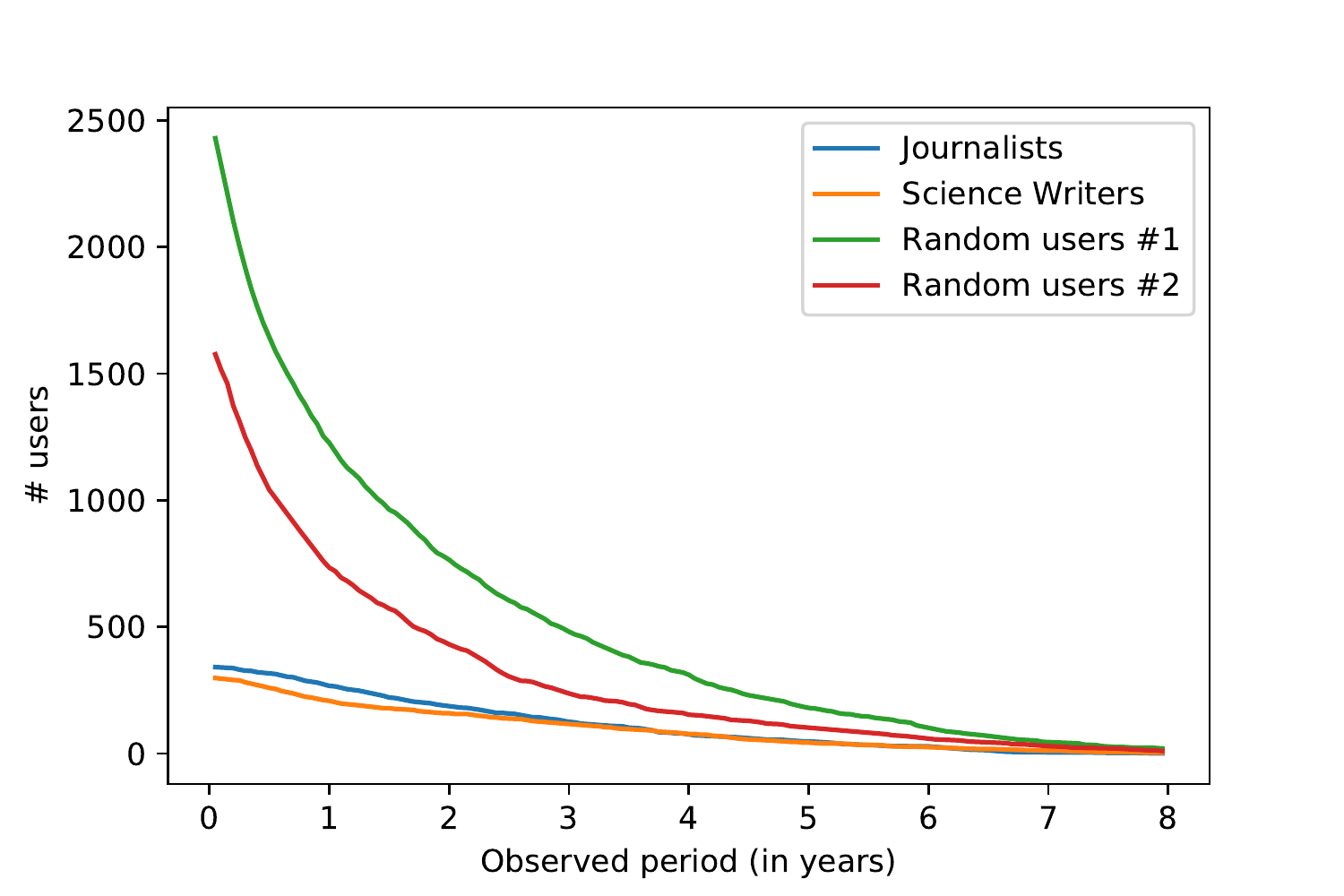}\vspace{-8pt}
  \captionof{figure}{Number of selected timelines depending on the observation window.}
  \label{fig:nb-users}
\end{minipage} %
\hspace{5pt}
\begin{minipage}{.48\textwidth}
  \centering
  \includegraphics[width=\linewidth]{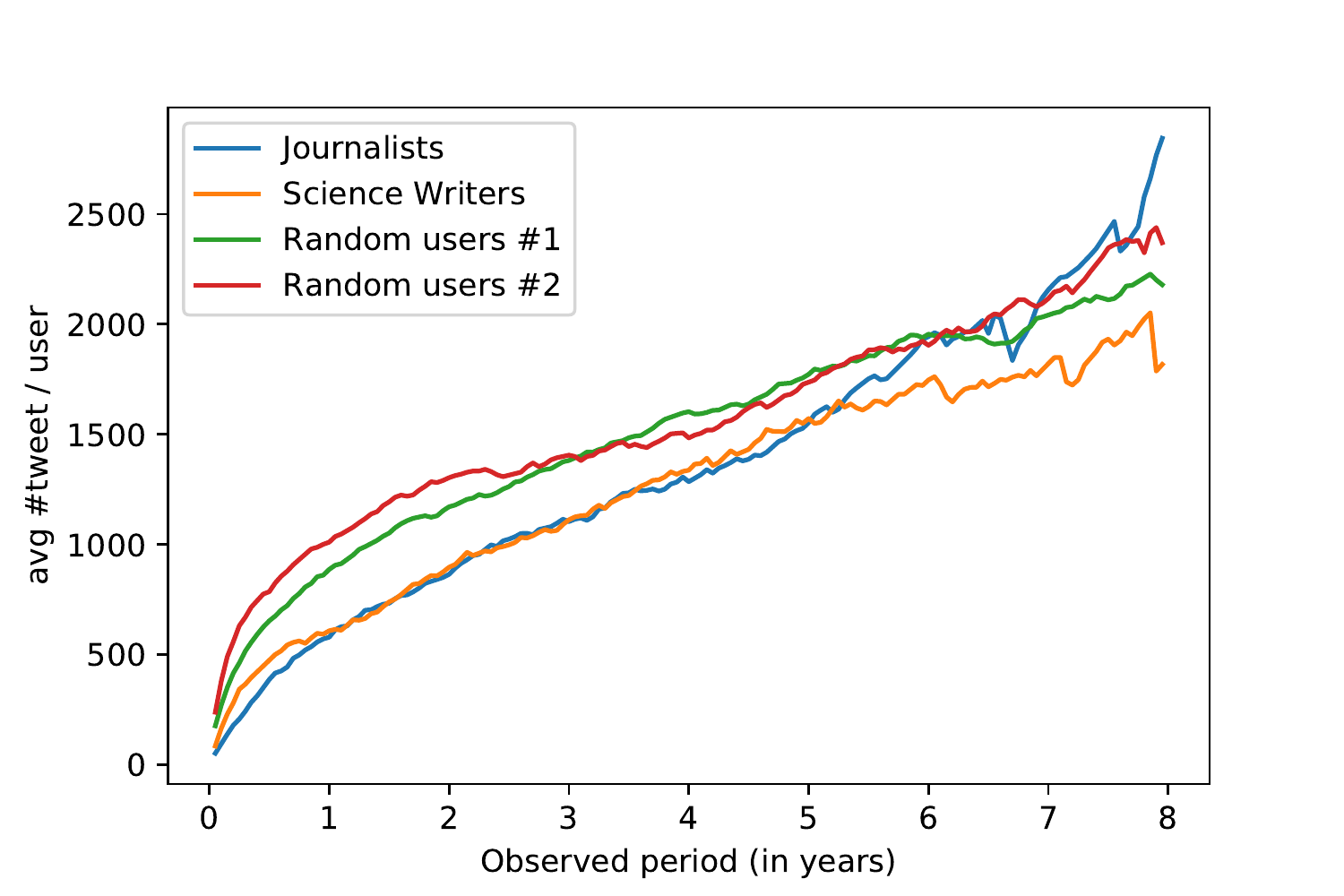}\vspace{-8pt}
  \captionof{figure}{Average number of tweets depending on the observation window.}
  \label{fig:avg-tweet}
\end{minipage}\vspace{-10pt}
\end{figure}

The observed timeline size is only constrained by the number of tweets (limited by API), thus the observation period varies according to the frequency with which the account is tweeting: for very active users, the last 3200 tweets will only cover a short time span. This raises the following problem: as random users are generally more active, their observation period is shorter
, and this may create a significant sampling bias. In fact, the length of the observation period affects the measured word usage frequencies discussed in Section~\ref{sec:preliminaries} (specifically, we cannot observe frequencies lower than the inverse of the observation period). In order to guarantee a fair comparison across user categories and to be able to compare users with different tweeting activities without introducing biases, we choose to work on timelines with the same duration, by restricting to an observation window $T$. To obtain timelines that have the same observation window $T$ (in years), we delete all those with a duration shorter than $T$ and remove tweets written more than $T$ years ago from the remaining ones. Increasing $T$ therefore reduces the number of profiles we can keep (see Figure \ref{fig:nb-users}): for a $T$ larger than 2 years, that number is divided by two, and for a $T$ larger than 3 years, it falls below 500 for all datasets. On the contrary, the average number of tweets per timeline increases linearly with $T$ (Figure \ref{fig:avg-tweet}). The choice of an observation window will then result from a trade-off between a high number of timelines per dataset and a large average number of tweets per timeline. To simplify the choice of $T$, we only select round numbers of years. We can read in Table \ref{tab:volume-same-period} that, beyond 3 years, the number of users falls below 100 for some datasets. On the other hand, the number of tweets for $T = 1 \textrm{ year}$ remains acceptable ($> 500$). We, therefore, decided to carry on the analysis with $T \in \{1\textrm{ year}, 2\textrm{ years}, 3 \textrm{ years}\}$. Please note that random users have a higher frequency of tweeting than others. This difference tends to smooth out when the observation period is longer (Table \ref{tab:volume-same-period}). This can be explained by the fact that the timelines with the highest tweet frequency are excluded in that case because their observation period is too small.

\begin{table}[t]
\scriptsize
\center
\setlength{\tabcolsep}{0.5em}
\renewcommand{\arraystretch}{1.2}
    \begin{tabular}{lccccccc}
        \toprule
        \multirow{2}{*}{Datasets} & \multicolumn{4}{c}{Number of users} & \multicolumn{3}{c}{Avg \# of tweets / user} \\ \cline{2-8} 
        & 1 year & 2 years & 3 years &  4 years & 1 year & 2 years & 3 years \\ \midrule
            NYT Journalists & 268 & 187  & 125 & 75 & 579.71 & 865.02 & 1104.58 \\
            Science Writers & 208 & 159 & 117 & 77 & 609.08 & 897.29 & 1112.63\\
            Random Users \#1 & 1227 & 765 & 481 & 311 & 897.29 & 1179.98 & 1403.50\\
            Random Users \#2 & 734 & 431 & 237 & 153 & 1057.41 & 1315.71 & 1404.60\\
        \bottomrule
    \end{tabular}\vspace{-15pt}
    \caption{Number of users and tweeting frequency at different observation windows.}
    \label{tab:volume-same-period}\vspace{-30pt}
\end{table}

\vspace{-10pt}
\subsection{Word extraction}
\vspace{-8pt}

Since the analysis has a focus on words and their frequency of use, we take advantage of NLP techniques for extracting them. As first step, all the syntactic marks that are specific to communication in online social networks (mentions with~@, hashtags with \#, links, emojis) are discarded (see Table~\ref{tab:removedtokens} in Appendix~\ref{app:additionaltables} for a summary). 
Once the remaining words are tokenized (i.e., identified as words), those that are used to articulate the sentence (e.g., ``with", ``a", ``but") are dropped. This type of words is called a functional word as opposed (in linguistics) to lexical words, which have a meaning independent of the context. These two categories involve different cognitive processes (syntactic for functional words and semantic for lexical words), different parts of the brain~\cite{diaz2009comparison}, and probably different neurological organizations~\cite{friederici2000segregating}. We are more interested in lexical words because their frequency in written production depends on the author's intentions, as opposed to functional words frequencies that depend on the language characteristics\footnote{Functional words may also depend on the style of an author (and due to this they are often used in stylometry). Still, whether their usage require a significant cognitive effort is arguable, hence in this work we opted for their removal.}. Moreover, lexical words represent the biggest part of the vocabulary. Functional words are generally called stop-words in the NLP domain and many libraries provide tools to filter them out. 

As this work will leverage word frequencies as a proxy for discovering cognitive properties, we need to group words derived from the same root (e.g. ``work" and ``worked") in order to calculate their number of occurrences. This operation can be achieved with two methods: stemming and lemmatization. Stemming algorithms generally remove the last letters thanks to complex heuristics, whereas lemmatization uses the dictionary and a real morphological analysis of the word to find its normalized form. Stemming is faster, but it may cause some mistakes of overstemming and understemming. For this reason, we choose to perform lemmatization. 
Once we have obtained the number of occurrences for each word base, we remove all those that appear only once to leave out the majority of misspelled words.
Table \ref{tab:token-lemma} in Appendix~\ref{app:additionaltables} contains examples of the entire preprocessing part.

%% file: body.tex
\vspace{-5pt}

Recalling that our goal is to investigate the structure and structural invariants in language production, in this section we present the main findings of our study. The methodology of our analysis is as follows. First, we analyse the frequency of words usage and the richness in vocabulary across the datasets, as preliminary characterisation of the different types of users we consider (Section~\ref{sec:preliminaries}). We then perform a clustering analysis to investigate whether regular groups of words used at different frequencies can be identified. This is the same method used to analyse the structural properties of human social networks, as explained in Section~\ref{sec:intro}. Based on this analysis, we describe the structural invariants that we found in Sections~\ref{sec:groups} and \ref{sec:group-sizes}.

\vspace{-15pt}
\subsection{Preliminaries}
\label{sec:preliminaries}
\vspace{-5pt}

Let us focus on a tagged user $j$. When studying the social cognitive constraints, the contact frequency between two people was taken as proxy for their intimacy and, as a result, for their cognitive effort in nurturing the relationship. Similarly, the frequency $f_i$ at which user $j$ uses word $i$ is considered here as a proxy of their ``relationship". Frequency $f_i$ is given by $\frac{n_{ij}}{t_i}$, where $n_{ij}$ denotes the number of occurrences of word $i$ in user $j$'s timeline, and $t_i$ denotes the observation window of $j$'s account in years. 

\begin{figure}[t]
  \centering
  \includegraphics[width=0.8\linewidth]{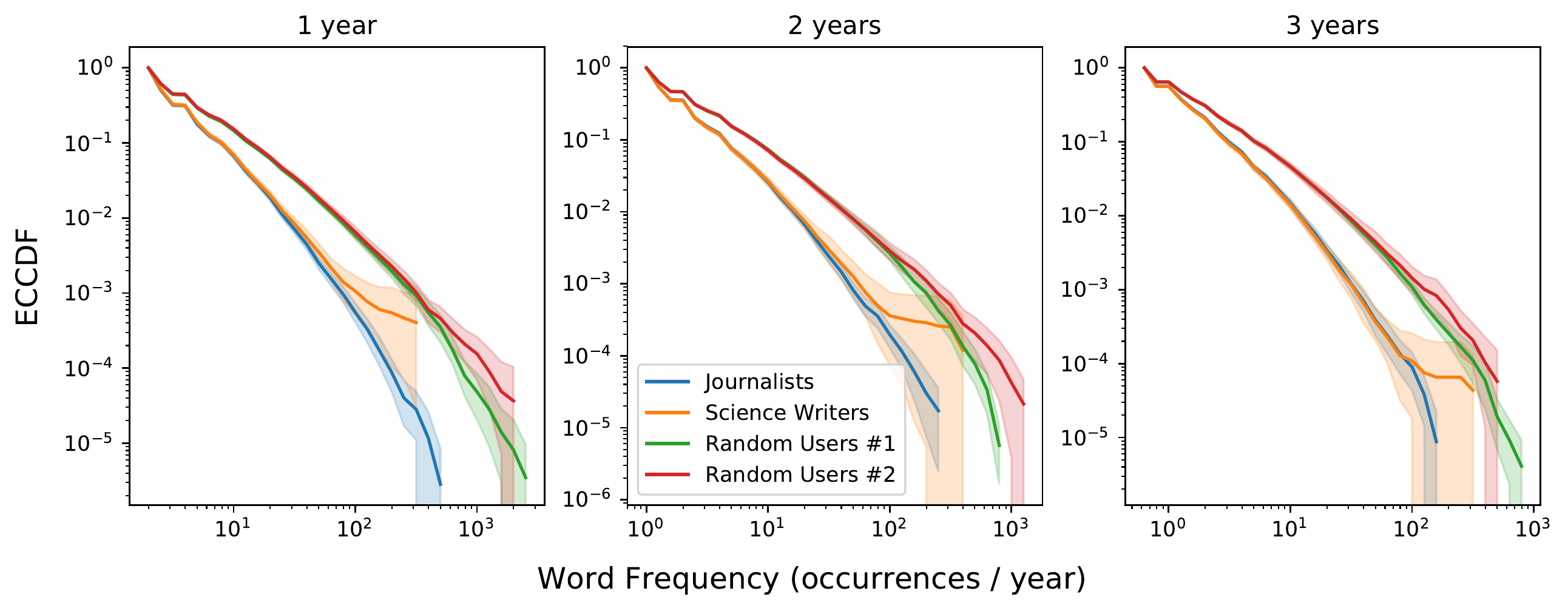}\vspace{-10pt}
  \caption{Aggregate visualization of user-wise empirical CCDF of word usage frequencies, in log-log scale. The solid line corresponds to the average and the shaded area to the 95\% confidence intervals.}
  \label{fig:log-freq}\vspace{-20pt}
\end{figure}

\begin{figure}[t]
\centering
\begin{minipage}{.48\textwidth}
  \centering
  \includegraphics[width=0.8\linewidth]{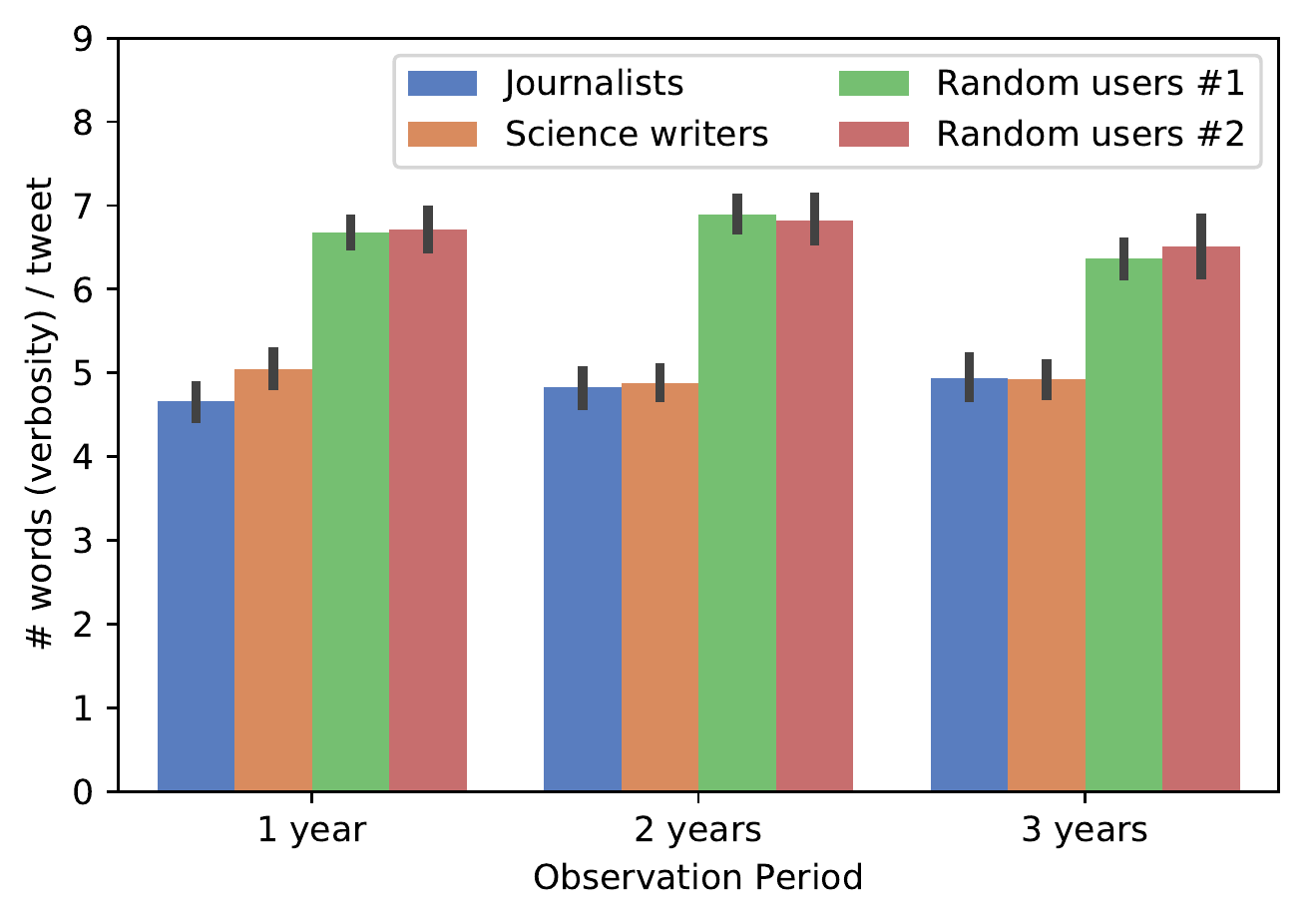}\vspace{-5pt}
        \captionof{figure}{Average verbosity, with 95\% confidence intervals.}
        \label{fig:verbosity}
\end{minipage}%
\hspace{5pt}
\begin{minipage}{.48\textwidth}
  \centering
        \includegraphics[width=0.8\linewidth]{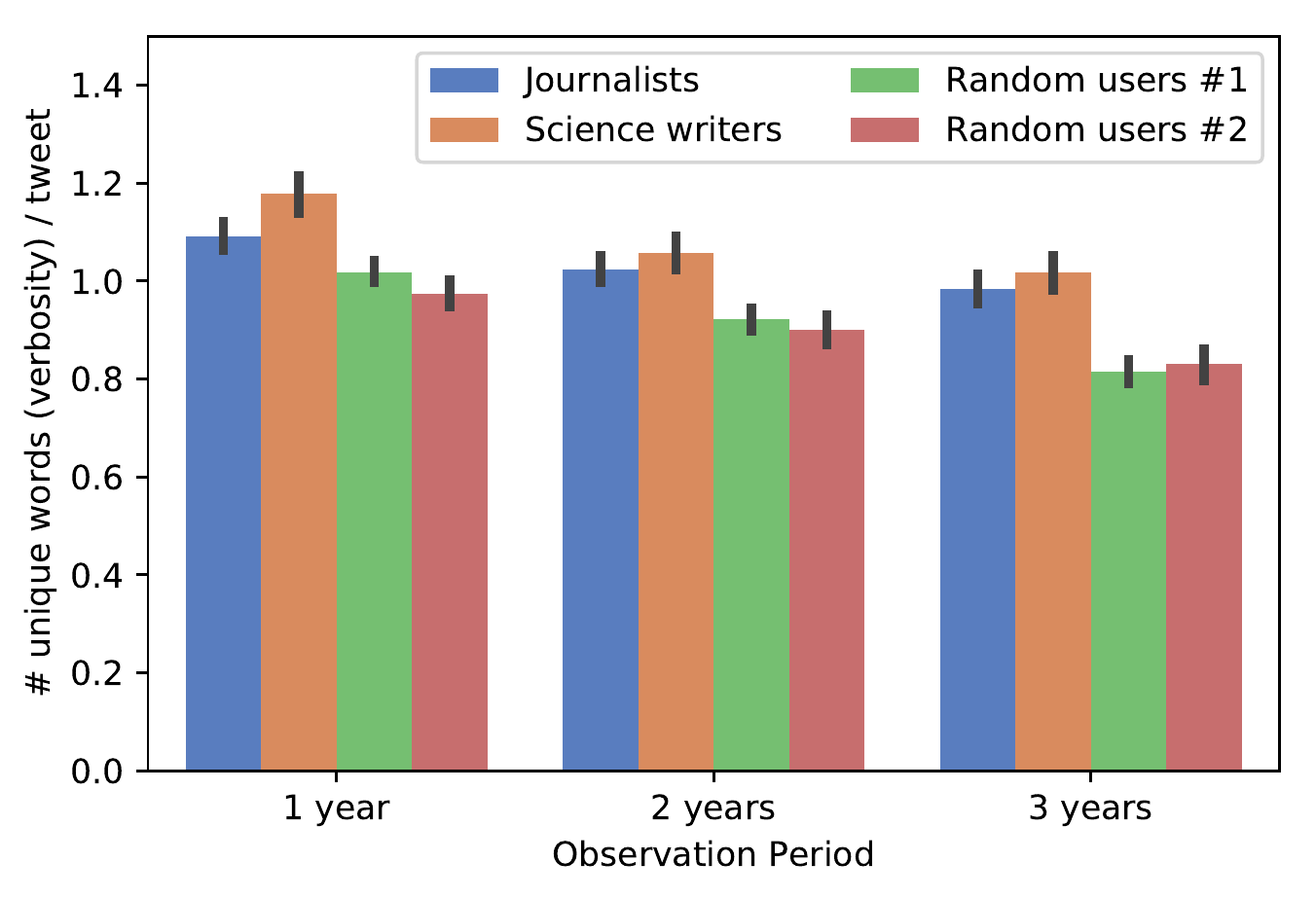}\vspace{-5pt}
        \captionof{figure}{Average lexical richness, with 95\% confidence intervals.}
        \label{fig:lexrichness}
        
\end{minipage}
\vspace{-20pt}
\end{figure}

Figure~\ref{fig:log-freq} shows the frequency distribution for the different categories of users (regular users vs professional writers) and for the different observation windows (1, 2, 3 years). We can make two observations. First, the distributions exhibit a heavy-tailed behaviour (see Table~\ref{tab:pareto-fit} in Appendix~\ref{app:additionaltables} for the fitting of individual users word frequency). 
Second, the distributions are very similar two by two: specialized users (journalists and science writers) who fulfill a particular role of information in the social network, and randoms users who are samples of more regular users. The first group seems to use more low-frequency words, while the second group uses a larger proportion of high-frequency words.
Based on the second observation, we can compare the datasets based on two criteria: \emph{verbosity}, which counts the total number of words per tweet, and \emph{lexical richness}, which counts the number of distinct words per tweet (Figures~\ref{fig:verbosity}-\ref{fig:lexrichness}). Despite  a lower verbosity (Figure \ref{fig:verbosity}), the vocabulary of specialized users seems richer than those of random users (Figure \ref{fig:lexrichness}). This is intuitive, but not necessarily expected. Professional users certainly have a higher diversity in the use of words. However, this manifests also in Twitter, i.e., in an environment where they do not write necessarily for professional reasons, but where they are (supposedly) writing in a more immediate and informal way.

\vspace{-10pt}
\subsection{Many words, just a few groups}
\label{sec:groups}
\vspace{-5pt}

Using the frequencies described in the previous section, we now investigate whether the words of a user can be grouped into homogeneous classes, and whether different users feature a similar number of classes or not. To this aim, for each user, we leverage a clustering algorithm to group words with a similar frequency. The selected algorithm is Mean Shift~\cite{fukunaga1975estimation}, because as opposed to Jenks~\cite{jenks1977optimal} or KMeans~\cite{macqueen1967some}, it is able to find the optimal number of clusters without fixed parameters. The original Mean Shift algorithm has nevertheless a drawback: it is only able to find the estimated density peaks with a fixed bandwidth kernel. The bandwidth is estimated based on the distribution of the pairwise distances between all the frequency values. However, in our case the distance between frequencies is not homogeneous: most of them are concentrated in the lowest values, close to each other. Hence, the selected bandwidth is fitted for estimating the density in that area, but not in the tail of the distribution. 
For that reason, a log-transformation is applied to the frequency values prior to the Mean Shift run: it still allows a fine mode detection in low-frequency part and compresses high values to allow detection of modes with a larger width. The use of a logarithmic scale is also used by psychological researchers to explain the impact of word frequency on their cognitive processing~\cite{brysbaert2018word}. 

\begin{figure}[t]
  \centering
  \includegraphics[width=\linewidth]{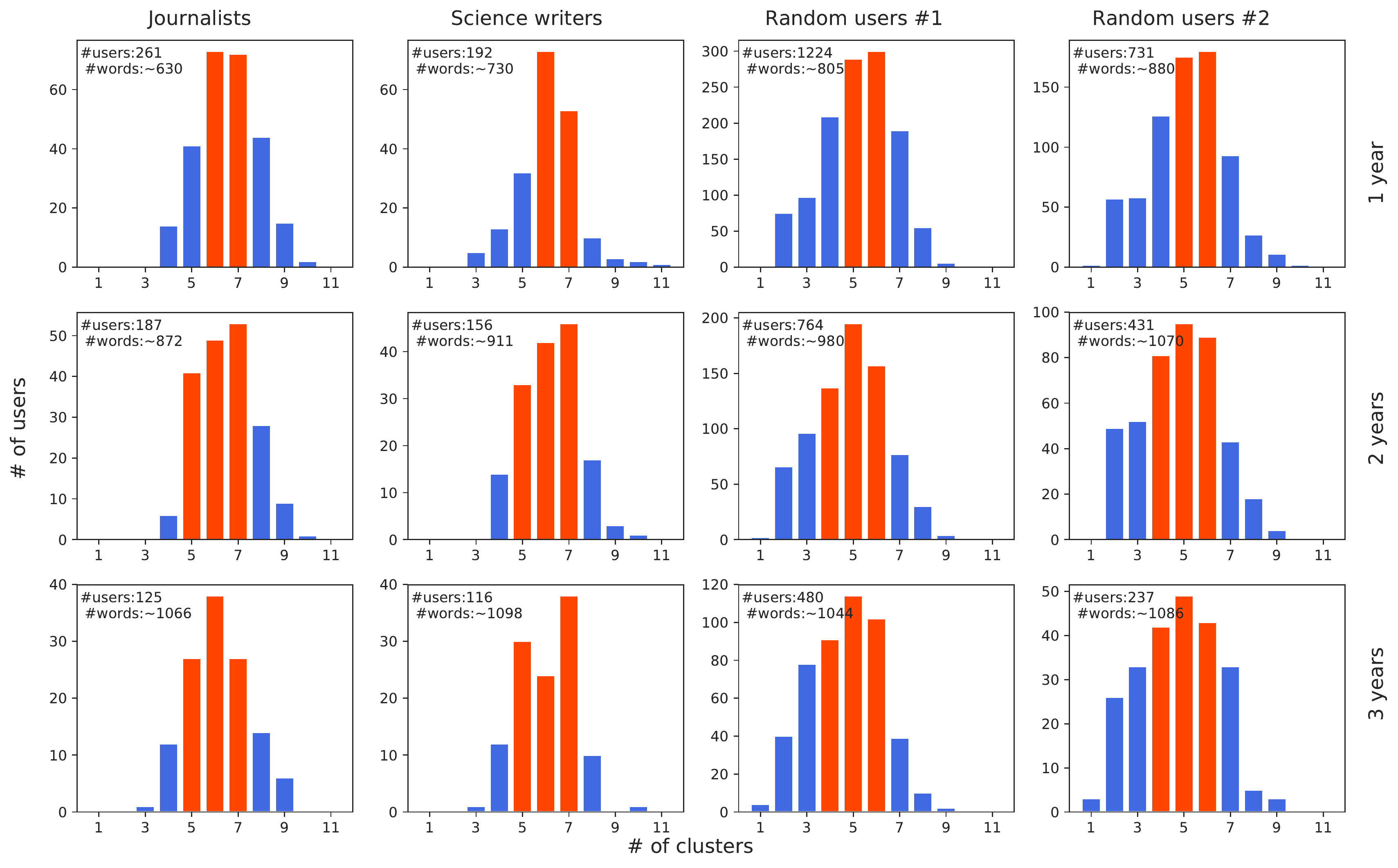}
  \caption{Number of clusters obtained applying Mean Shift to log-transformed frequencies. The most frequent number of clusters are highlighted in red. 
  \label{fig:clusters-hist}}
  \vspace{-20pt}
\end{figure}

The histograms of the obtained optimal number of clusters are shown in Figure~\ref{fig:clusters-hist}. It is interesting to note that, despite the heterogeneity of users (in terms of tweeting frequency, verbosity, and lexical richness), the distributions are always quite narrow, with peaks consistently between 5 and 7 clusters. The observation period seems to have a very limited effect on the resulting cluster structure. This means that, after one year, the different groups of words can be already identified reliably. In addition, this limited effect actually reinforces the idea of a natural grouping: when more words are added (longer observation period) the clusters become slightly fewer, not slightly more. Hence, new words tend to reinforce existing clusters. Thus, similarly to the social constraints case, also for language production we observe a fairly regular and consistent structure. This is the first important result of the paper, hinting at the existence of structural invariants in cognitive processes, which we summarise below.

\vspace{-7pt}
\begin{description}
    \item[Cognitive constraint 1:] Individual distributions of word frequencies are divided into a consistent number of groups. Since word frequencies impact the cognitive processes underlying word learning and retrieval in the mental lexicon~\cite{perfetti2005word}, these groups can be an indirect trace of these processes' properties. The number of groups is only marginally affected by the class (specialized or generic) the users belong to or by the observation window. This regularity might also suggest that these groups of words correspond to linguistic functional groups, and we plan to investigate this as future work. 
\end{description}


\vspace{-15pt}
\subsection{Exploring the group sizes \label{sec:group-sizes}}
\vspace{-5pt}

We now study the size of the clusters identified in the previous section. For the sake of statistical reliability, we only consider those users whose optimal number of clusters (as identified by Mean Shift) corresponds to the most popular number of clusters (red bars) in Figure~\ref{fig:clusters-hist}. This allows us to have a sufficient number of samples in each class. 
We rank each cluster by its position in the frequency distribution: cluster \#1 is the one that contains the most frequent words, and the last cluster is the one that contains the least used. Following the convention of the Dunbar's model discussed in Section~\ref{sec:intro}, these clusters can be mapped into concentric layers (or circles), which provide a cumulative view of word usage. Specifically, layer $i$ includes all clusters from the first to the $i$-th. Layers provide a convenient grouping of words used \emph{at least} at a certain frequency. We refer to this layered structure as \emph{ego network of words}.  

\begin{figure}[t]
  \centering
  \includegraphics[width=\linewidth]{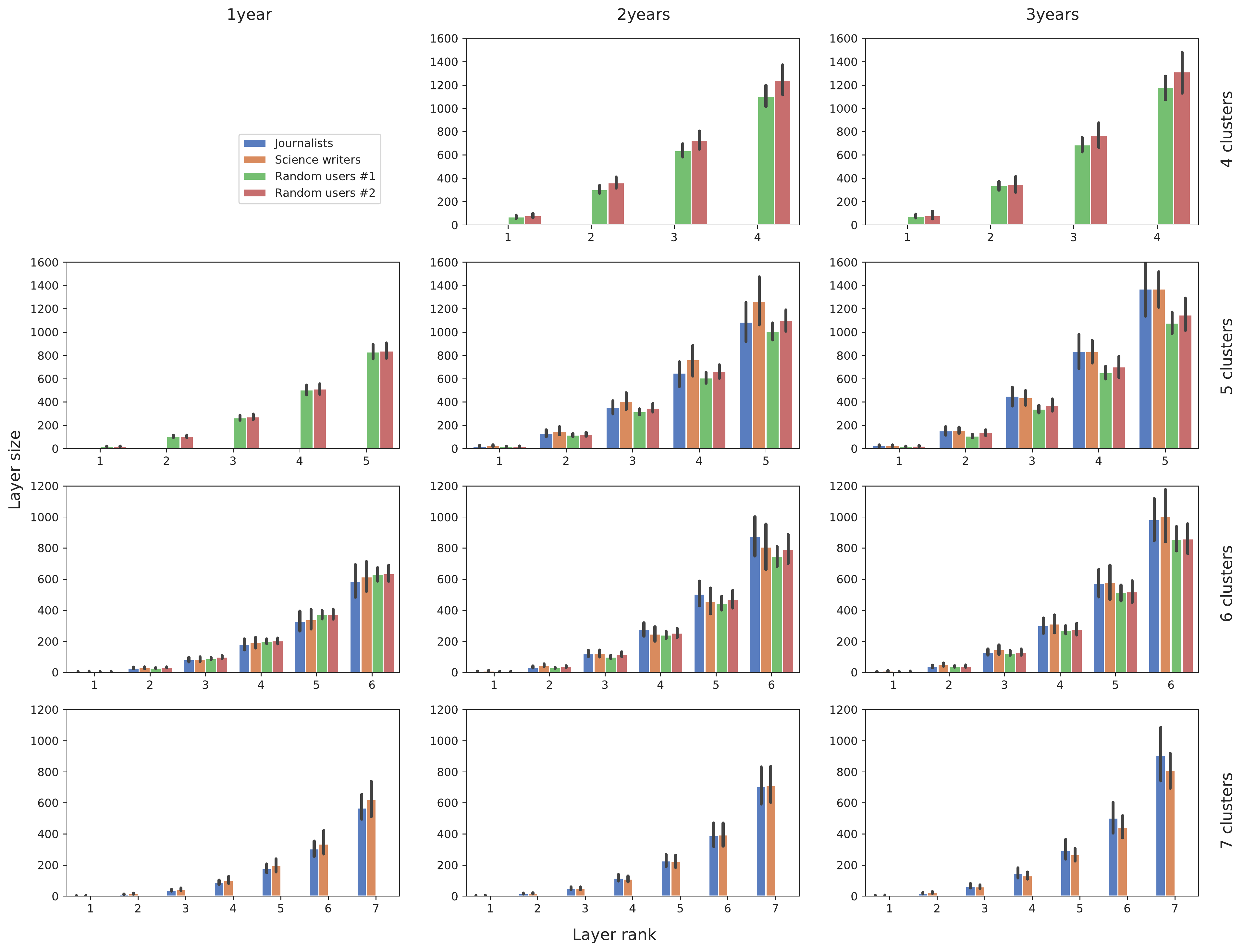}\vspace{-10pt}
  \caption{Average layer size (with 95\% confidence intervals) for the various datasets,  different number of clusters (rows), and different observation periods (columns).}
  \label{fig:cumul-size}\vspace{-20pt}
\end{figure}

Figure~\ref{fig:cumul-size} shows the average layer sizes for every dataset and different observation periods. As expected, for a given  number of clusters, the layer size increases as we expand the observation period, because more words are brought in. For a given number of clusters  we also observe a striking regularity across the datasets, with confidence intervals overlapping in practically all settings. Typically, the layer sizes are slightly higher for journalists and science writers ($T=2$ years and $T=3$ years). The main reason is that their lexicon is generally richer than those of regular users (as discussed in Sec.~\ref{sec:preliminaries}) and this is reflected in their layer size.



\begin{figure}[p]
  \centering
  \includegraphics[width=\linewidth]{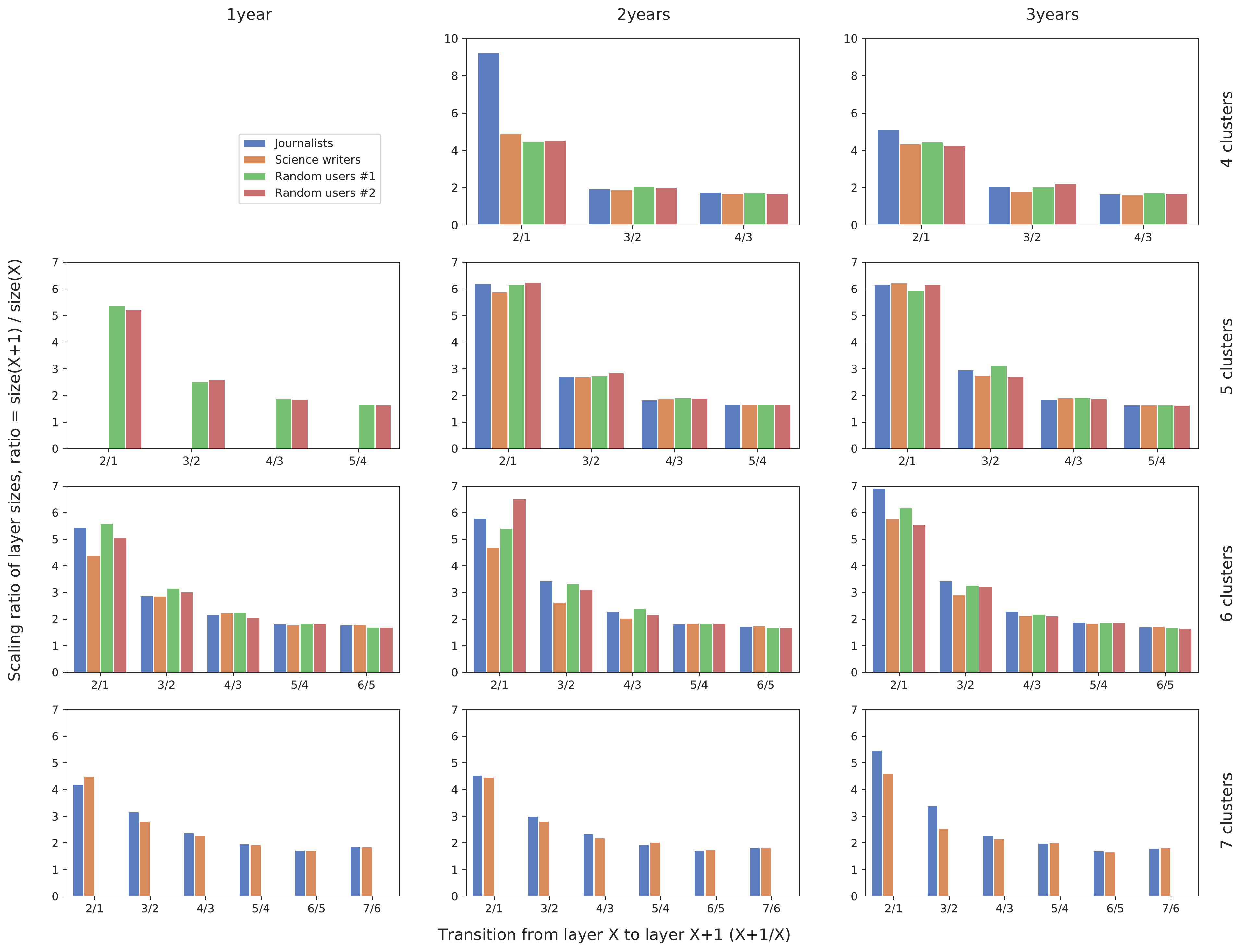}\vspace{-10pt}
  \caption{Scaling ratio for the various datasets, different number of clusters (rows), and different observation periods (columns).}
  \label{fig:scaling-ratio}\vspace{-10pt}
\end{figure}

Another typical metric that is analysed in the context of social cognitive constraints is the scaling ratio between layers, which, as discussed earlier, corresponds to the ratio between the size of consecutive layers. The scaling ratio is an important measure of regularity, as it captures a relative pattern across layers, beyond the absolute values of their size. Figure~\ref{fig:scaling-ratio} shows the scaling ratio of the layers in language production. We can observe the following general behavior: the scaling ratio starts with a high value between layers \#1 and \#2 , but always gets closer to 2-3 as we move outwards. This empirical rule is valid whatever the dataset and whatever the observation period. This is another significant structural regularity, quite similar to the one found for social ego networks, as a further hint of cognitive constraints behind the way humans organise word use.

\begin{figure}[p]
  \centering
  \includegraphics[width=\linewidth]{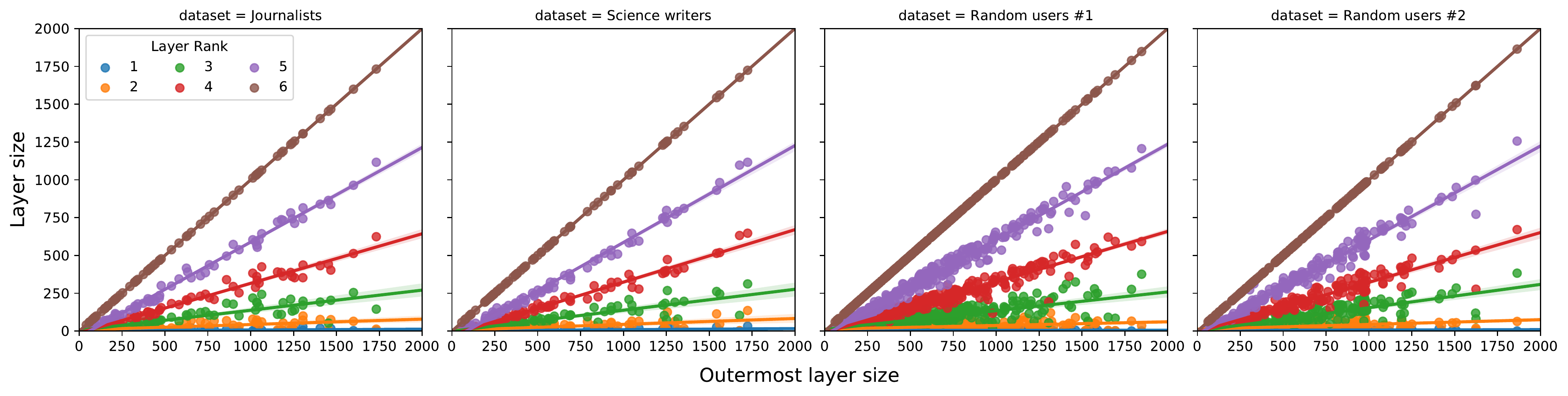}\vspace{-10pt}
  \caption{Linear regression between the total number of unique words used by each user (corresponding to the size of the outermost layer) and the individual layer sizes.}
  \label{fig:correl-cumul-size}
\end{figure}

In order to further investigate the structure of the word clusters, we compute the linear regression coefficients between the total number of unique words used by each user (corresponding to the size of the outermost layer) and the individual layer sizes. Due to space limits, in Table~\ref{table:coeff} we only report the exact coefficients for the journalists dataset with T=1 year (but analogous results are obtained for the other categories and observation windows) and in Figure~\ref{fig:correl-cumul-size} we plot the linear regression for all the user categories with $T=1$ year. Note that the size of the most external cluster is basically the total number of words used by an individual in the observation window. It is thus interesting to see what happens when this number increases, i.e., if users who use more words distribute them uniformly across the clusters, or not. Table~\ref{table:coeff} shows two interesting features. First, it shows another regularity, as the size of all layers linearly increases with the most external cluster size, with the exception of the first one (Figure~\ref{fig:correl-cumul-size}). Moreover, it is quite interesting to observe that the second-last and third-last layer consistently account for approximately 60\% and 30\% of the used words, irrespective of the number of clusters. This indicates that users with more clusters split at a finer granularity words used at highest frequencies, i.e., they organise differently their innermost clusters, without modifying significantly the size of the most external ones.

As a final comment on Fig.~\ref{fig:scaling-ratio}, please note that the innermost layer tends to be approximately five times smaller than the next one. This suggests that this layer, containing the most used words, may be drastically different from the others (as also evident from Table~\ref{table:coeff}).
We leave as future work the characterization of this special layer and we summarise below the main  results of the section.

\vspace{-5pt}
\begin{description}
    \item[Cognitive constraint 2:] Structural invariants in terms of layer sizes and scaling ratio are observed also in the language domain. Specifically, we found that the size of the layers approximately doubles when moving from layer $i$ to layer $i + 1$, with the only exception of the first layer.
    \item[Cognitive constraint 3:] Users with more clusters organise differently their innermost clusters, without modifying significantly the size of the most external ones, which consistently account for approximately 60\% and 30\% of the used words, irrespective of the number of clusters of the user. 
\end{description}

\begin{table}[p]
\scriptsize
\center
\begin{tabular}{c c c c c c c c }
\toprule
\multirow{2}{*}{Opt. \# of clusters  } & \multicolumn{7}{c}{Cluster Rank}           \\ \cmidrule(r){2-8} 
                                    & 1    & 2    & 3    & 4    & 5    & 6    & 7 \\ \midrule
4 clusters                          & 0.04 & \cellcolor{gray!25} 0.33 & \cellcolor{gray!50} 0.61 & 1.00    &      &      &   \\ 
5 clusters                          & 0.02 & \cellcolor{gray!10}0.13 & \cellcolor{gray!25} 0.33 & \cellcolor{gray!50} 0.62 & 1.00    &      &   \\ 
6 clusters                          & 0.01 & 0.04 & \cellcolor{gray!10}0.14 & \cellcolor{gray!25} 0.32 & \cellcolor{gray!50} 0.59 & 1.00    &   \\ 
7 clusters                          & 0.00 & 0.02 & 0.06 & \cellcolor{gray!10}0.16 & \cellcolor{gray!25} 0.32 & \cellcolor{gray!50} 0.56 & 1.00 \\ \bottomrule
\end{tabular}\vspace{-15pt}
\caption{Linear coefficients obtained for the journalists dataset with $T=1$ year. \label{table:coeff}}\vspace{-25pt}
\end{table}

%% file: conclusion.tex
\vspace{-8pt}

In this paper, we investigated, through a data-driven approach, whether a regular structure can be found in the way people use words, as a symptom of cognitive constraints in their mental process. 
This is motivated by the fact that other mental processes are known to be driven by cognitive constraints, such as the way how humans allocate cognitive capacity to social relationships. 
To this aim, we collected a diverse dataset from Twitter (identified as one of the major sources of informal and spontaneous language online), including tweets from regular Twitter users and from professional writers. 
Then, leveraging a methodology similar to the one used to uncover social constraints, we have analysed the structural properties of language production on Twitter, uncovering regularities that constitute preliminary evidence of the aforementioned cognitive constraints.
Specifically, we have found that, similarly to the social case, a concentric layered structure (ego network of words) very well captures how an individual organizes their cognitive effort in language production. 
Words can be grouped typically in between 5 and 7 layers, regardless of the specific class of users. We also observe a structural invariant in the size of the layers, which grow approximately 2-3 times when moving from a layer to the next one. 
A second structural invariant emerges for the external layers, which, regardless of the number of clusters of the user, consistently account for approximately 60\% and 30\% of the used words.
While these findings are restricted to language production on Twitter, as future work we plan to generalize these results to other types of written communication.

%% file: acks.tex

This work was partially funded by the SoBigData++, HumaneAI-Net, MARVEL, and OK-INSAID projects. The SoBigData++ project has received funding from the European Union's Horizon 2020 research and innovation programme under grant agreement No 871042. The HumaneAI-Net project has received funding from the European Union's Horizon 2020 research and innovation programme under grant agreement No 952026. The MARVEL project has received funding from the European Union's Horizon 2020 research and innovation programme under grant agreement No 957337. The OK-INSAID project has received funding from the Italian PON-MISE program under grant agreement ARS01 00917.

%% file: appendix.tex
\subsection{Identifying active Twitter users}
\label{app:activeusers}

In order to be relevant to our work, a Twitter account must be an active account, which we define as an account not abandoned by its user and that tweets regularly. A Twitter account is considered abandoned, and we discard it, if the time since the last tweet is significantly bigger (we set this threshold at 6 months, as previously done also in~\cite{boldrini2018twitter}) than the largest period of inactivity for the account. We also consider the tweeting regularity, measured by counting the number of months where the user has been inactive. The account is tagged as sporadic, and discarded, if this number of months represents more than 50\% of the observation period (defined as the time between the first tweet of a user in our dataset and the download time). We also discard accounts whose entire timeline is covered by the 3200 tweets that we are able to download, because their Twitter behaviour might have yet to stabilise (it is known that the tweeting activity needs a few months after an account is created to stabilise). 

\subsection{Additional tables}
\label{app:additionaltables}


\begin{table}[H]
\center
\setlength{\tabcolsep}{0.5em}
\renewcommand{\arraystretch}{1.2}
\begin{tabular}{lccc}
\toprule
                 & Percentage of hashtags & Percentage of links & Percentage of emojis \\
\midrule
Journalists      & 1.34 \%           & 7.27 \%        & 0.20 \%         \\
Science writers  & 3.47 \%           & 8.02 \%        & 0.55 \%         \\
Random users \#1 & 16.84 \%          & 6.97 \%        & 5.21 \%         \\
Random users \#2 & 7.20 \%           & 6.42 \%        & 4.60 \%        \\
\bottomrule
\end{tabular}
\caption{In the process of word extraction, the tweet is decomposed in tokens which are usually separated by spaces. These tokens generally corresponds to words, but they can also be links, emojis and others markers that are specific to the online language such as hashtags. The table gives the percentage of hashtags, links and emojis, which are tokens filtered out from the datasets.}
\label{tab:removedtokens}
\end{table}




\begin{table}[H]
    \scriptsize
    \setlength{\tabcolsep}{0.6em}
    \renewcommand{\arraystretch}{1.5}
    \begin{tabularx}{\textwidth}{XX}
        \toprule
        Original tweet content & List of words after pre-processing \\
        \midrule 
        The @Patriots say they don’t spy anymore. The @Eagles weren’t taking any chances. They ran a "fake" practice before the \#SuperBowl & spy, anymore, chance, run, fake, practice\\
        \#Paris attacks come 2 days before world leaders will meet in \#Turkey for the G20. Will be a huge test for Turkey. & attack, come, day, world, leader, meet, huge, test, turkey \\
        Latest garden species - the beautiful but destructive rosemary beetle, and a leafhopper (anyone know if this can be identified to species level from photo? Happy to give it a go)  \#30DaysWild \#MyWildCity \#gardening & late, garden, specie, beautiful, destructive, rosemary, beetle, leafhopper, know, identify, specie, level, photo, happy \\
        \bottomrule
    \end{tabularx}\vspace{-15pt}
    \caption{Example of word extraction results. \label{tab:token-lemma}}
    \vspace{-10pt}
\end{table}

\begin{table}[H]
    \scriptsize
    \renewcommand{\arraystretch}{1.5}
    \begin{tabular}{ccccccccccccc}
        \toprule
         & \multicolumn{3}{c}{Journalists} & \multicolumn{3}{c}{Science writers} & \multicolumn{3}{c}{Random dataset \#1} & \multicolumn{3}{c}{Random dataset \#2} \\
         \midrule
         & 1 year   & 2 years   & 3 years  & 1 year    & 2 years    & 3 years    & 1 year     & 2 years     & 3 years     & 1 year     & 2 years     & 3 years     \\
         \midrule
    p\textless{}0.1  & 22.1 \%   & 33.2 \%    & 49.6 \%   & 24.8 \%    & 37.3 \%     & 47.9 \%     & 41 \%       & 49.7 \%      & 52.6 \%      & 50.3 \%     & 62 \%        & 61.4 \%      \\
    p\textless{}0.05 & 17.6 \%   & 25.1 \%    & 39.2 \%   & 19.4 \%    & 29.1 \%     & 39.3 \%     & 32.8 \%     & 43.5 \%      & 43 \%        & 43.9 \%     & 52.1 \%      & 52.4 \%      \\
    p\textless{}0.01 & 8.6 \%    & 11.2 \%    & 18.4 \%   & 12.1 \%    & 15.8 \%     & 17.1 \%     & 21.4 \%     & 27.8 \%      & 27.8 \%      & 30.4 \%     & 37.3 \%      & 35.2 \%  \\
    \bottomrule
    \end{tabular}
    \caption{Percentage of users for which the hypothesis that the word frequency distribution is a power-law is rejected with a p-value below 0.1, 0.05 and 0.01. The p-value is obtained with the Kolmogorov–Smirnov test, using the fitting technique described in~\cite{clauset2009power}.
    \label{tab:pareto-fit}
    }
    \vspace{-10pt}
\end{table}